# A Mean-Field Description of Protein Folding: Evolutional Conservation of Hydrophobicity Profile and Folding Nuclei

S. NAKAMURA and O. NARIKIYO
*Department of Physics, Kyushu University, Fukuoka 810-8560, Japan*

**Abstract.** Evolutionally conserved quantity that specifies folding nuclei is pursued by a case study for a small protein (PDB code: 1ten). First it is demonstrated that the sequences of amino acids at folding nuclei are not conserved. Then 3D (3-dimensional) information of the structure is considered and it is found that a 3D hydrophobicity profile is essential to specify the folding nuclei and evolutionally conserved. This profile is maintained by the interaction, including entropic effect, among amino acids realized in the native state structure. Experimentally observed $\phi$-value is correlated to this 3D hydrophobicity profile after taking into account the effect of the contact distance in amino-acid sequence.

**Key word:** protein folding, folding nucleus, protein engineering, $\phi$-value, evolutional conservation, hydrophobic effect, contact distance

## 1. Introduction

In the process of protein folding the dynamics is heterogeneous and folding nuclei play a crucial role. Such folding nuclei are identified, for example, by $\phi$-value analysis in protein engineering experiments [1]. Recently it has been intensively discussed [2] whether the sequence of amino acids at the folding nucleus is evolutionally conserved. The conclusion seems to be negative [3]; the sequence is not evolutionally conserved. Then we try to find an evolutionally conserved quantity specifying folding nuclei in this paper.

As a case study we focus our attention to a small protein (PDB code: 1ten) whose folding nuclei are exhaustively studied [4]. First we check that the amino-acid sequence at the folding nucleus is not evolutionally conserved. Thus folding nuclei are not easily decided from 1D (1-dimensional) information of the sequence. Then we consider 3D (3-dimensional) information of the structure and find that a 3D-hydrophobicity profile (HP) is essential to specify the folding nuclei and evolutionally conserved. Experimentally observed $\phi$-value is not directly correlated to this 3D-HP, since the $\phi$-value is affected by the contact distance in amino-acid sequence. We can analyze the $\phi$-value taking this effect of the contact distance into account. Consequently we see that the folding nuclei are specified by the 3D-HP and this profile is maintained by the interaction among amino acids realized in the native state structure.

## 2. Non-Conservation of Sequence

After recent intensive discussions it has been concluded that the sequence of amino acids at the folding nucleus is not evolutionally conserved [5] on the basis of the evaluation of the entropy. We also check the non-conservation for our case of 1ten.

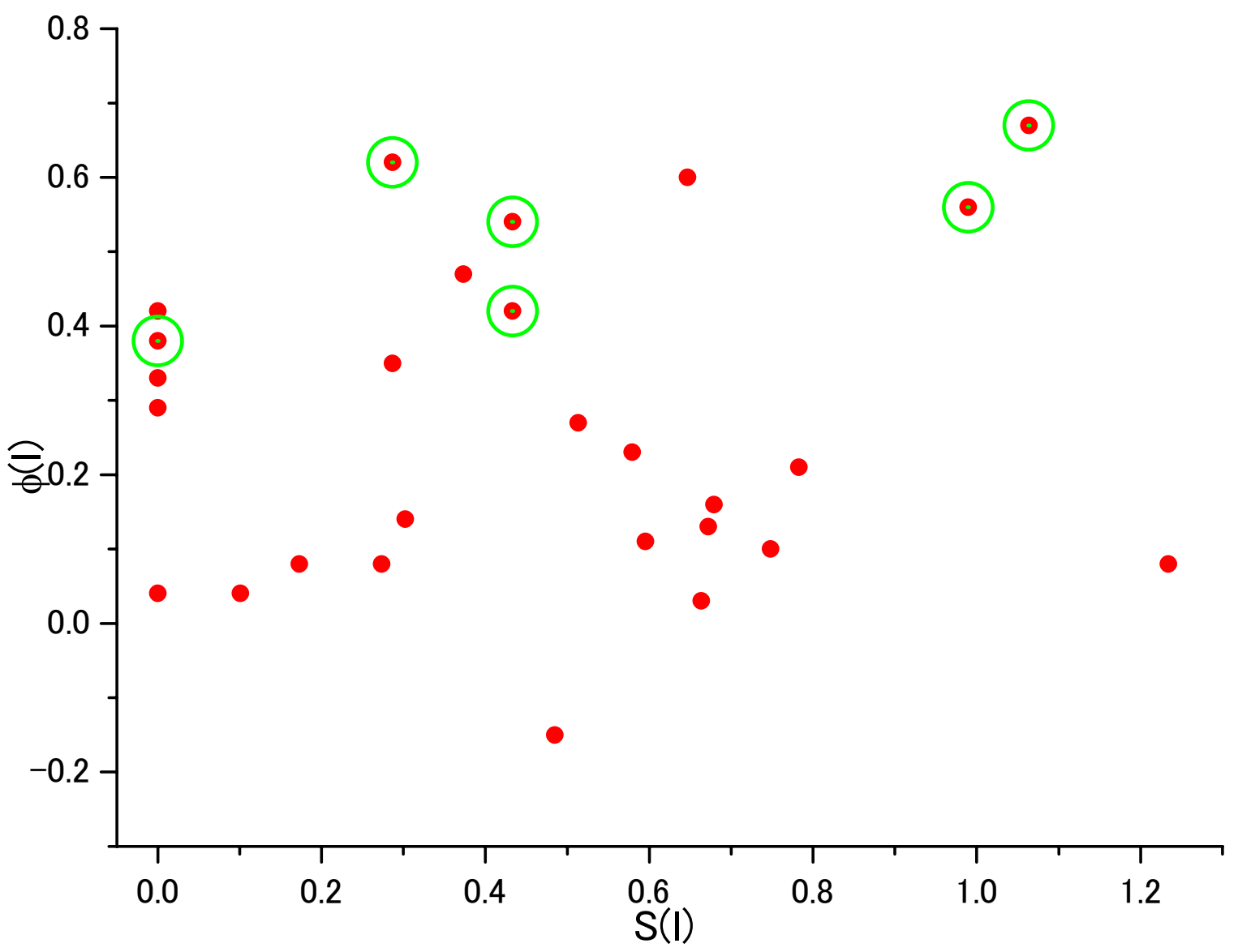

*Figure 1*. Correlation between the entropy *S*(*l*) and the $\phi$-value $\phi(l)$. The entropy for the *l*-th residue, *S*(*l*), is calculated as $S(l) = -\sum_i p_i(l) \log p_i(l)$ using the ensemble of 48 proteins resembling 1ten, obtained by HSSP in May 2005. Here $p_i(l)$ is the probability for finding an amino acid belonging to the *i*-th group in the *l*-th residue. The 6 groups are defined in the main text. The $\phi$-value at the *l*-th residue, $\phi(l)$, is taken from the experiment [4]. The data points for folding nuclei are circled.

We calculate the entropy of the amino-acid sequence by using 6-group scheme [6] where 20 kinds of amino acids are classified into 6 groups by their physico-chemical properties; {ALA , VAL , LEU , ILE , MET , CYS}, {PHE , TRP , TYR , HIS }, {SER , THR , ASN , GLN }, {LYS , ARG}, {ASP , GLU} and {GLY , PRO}. In this calculation amino acids in the same group are not distinguished so that the tendency for evolutional conservation is favored.

Hereafter we focus our attention to 6 folding nuclei [4], (the amino acid; the residue number in this paper, the residue number in PDB): (ILE; 20, 821), (TYR; 36, 837), (ILE; 48, 849), (LEU; 50, 851), (ILE; 59, 860) and (VAL; 70, 871), on secondary structures. Our mini-protein, 1ten, consists of 7 $\beta$-sheets.

In Figure 1 the correlation between the entropy and the $\phi$-value for 1ten is shown. It is seen that the sequence at the folding nucleus is not conserved.

## 3. Conservation of Hydrophobicity Profile

In the previous section we have shown that the amino-acid sequence at the folding nucleus is not conserved. Then we seek an evolutionally conserved quantity specifying folding nuclei, if it exists. Since folding nuclei are not easily decided from 1D information of the sequence, we consider 3D information of the structure.

A similar situation has been encountered in the study of the helix-turn-helix motif and it has been shown that 3D keynote [7], which consists of pairs of amino acids, can characterize the structure of the motif. The pairs are expected to have some correspondence to the folding nuclei. Thus we try to make the 3D keynote for 1ten. The pair is identified by the number of interactions between two amino acids. The interaction is assigned to each pairs of atoms, excluding hydrogen, if the diameter of the pair is smaller than 6 Å. In Figure 2 the number of interactions belonging to the amino acid is shown as a function of the sequence. The correlation between the number and the $\phi$-value is subtle so that the 3D keynote does not work for specifying the folding nuclei.

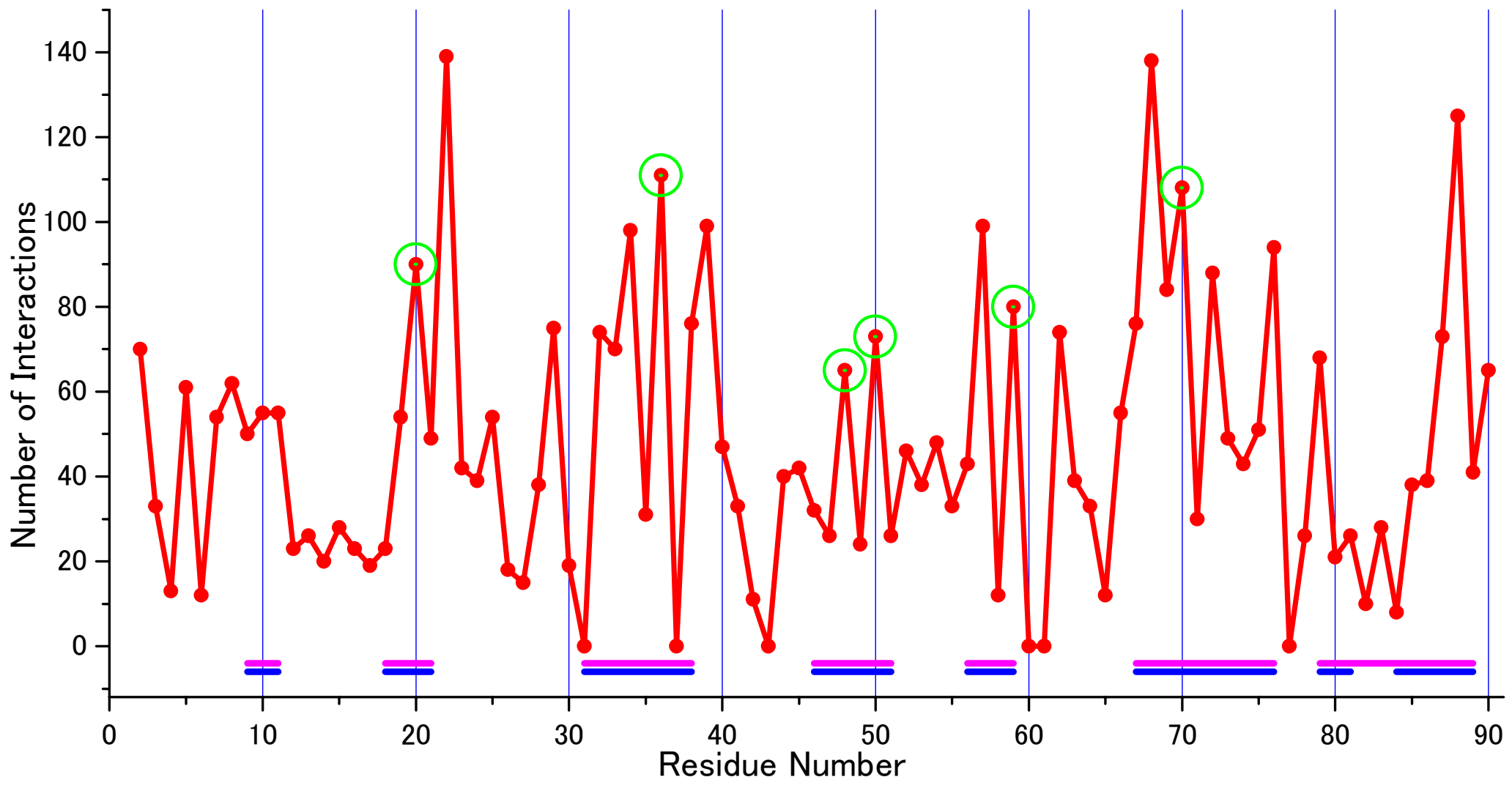

*Figure 2*. The number of interactions belonging to the amino acid, which is basic for the 3D key note, as a function of the sequence. The interaction is assigned to each pairs of atoms, excluding hydrogen, if the diameter of the pair is smaller than 6 Å. The protein 1ten consists of 7 $\beta$-sheets represented as the horizontal segments (upper) [4], while in PDB assignment 8 segments (lower) exist.

In the case of protein folding in general the formation of hydrophobic core plays an essential role. However, this entropic effect is not taken into account in the analysis of 3D keynote. Thus we introduce a 3D hydrophobicity in order to examine the hydrophobic effect. The 3D hydrophobicity for an amino acid is defined as the sum of the values of hydrophobicity [8] of amino acids within 12 Å from it. This 3D hydrophobicity reflects the structure of the native state. The structure is decided by interactions, including entropic effect, among amino acids. Here the folding nuclei are related to the formation of hydrophobic core. In Figure 3 the 3D-HP is shown as a function of the sequence. The 3D hydrophobicity and the $\phi$-value is correlated so that the 3D hydrophobicity can specify the folding nuclei. The correlation is further discussed in the next section.

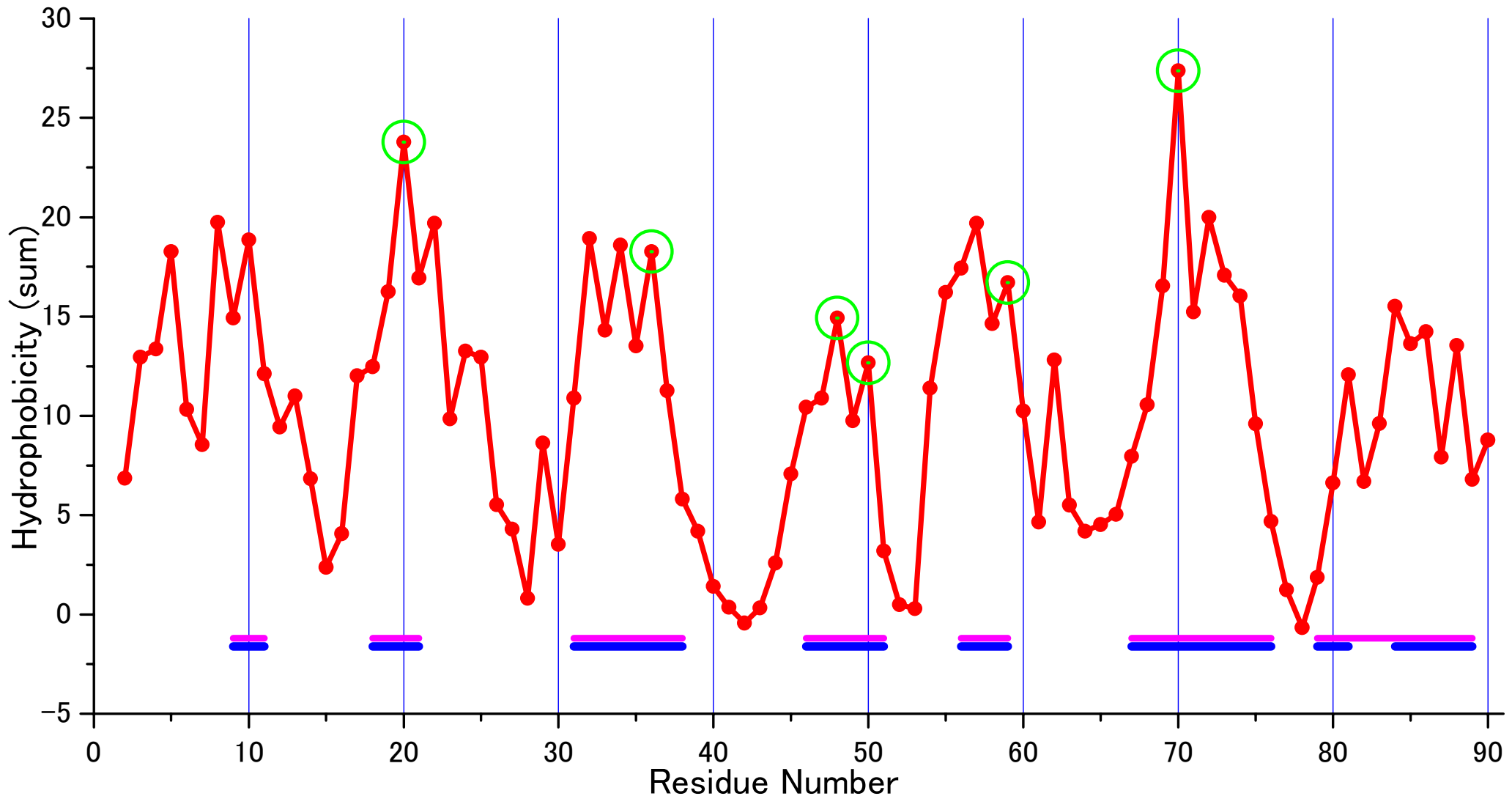

*Figure 3*. The 3D hydrophobicity profile as a function of amino-acid sequence. The 3D hydrophobicity for an amino acid is defined as the sum of the values of hydrophobicity of amino acids within 12 Å from it in the native state structure. The distance between amino acids is determined by that for $C_{\beta}$'s. In the case of GLY the position of $C_{\alpha}$ is employed exceptionally.

In order to see the evolutional property of the 3D-HP we plot several profiles with the highest Z-scores in HSSP compared with 1ten in Figure 4. From this plot it can be concluded that the 3D-HP is evolutionally conserved. In Figure 5 the correlation between the degree of conservation and the compactness determined by the distances among folding nuclei is shown. From this correlation it can be said that the structure in the native state is maintained by the 3D-HP. Here we have employed the compactness of the folding nuclei. On the other hand, it is hard to find key correlation shown in Figure 5 by the analysis of the whole structures of proteins, since the fragile loops are arranged to support the compactness but lead to the deviation of the structure.

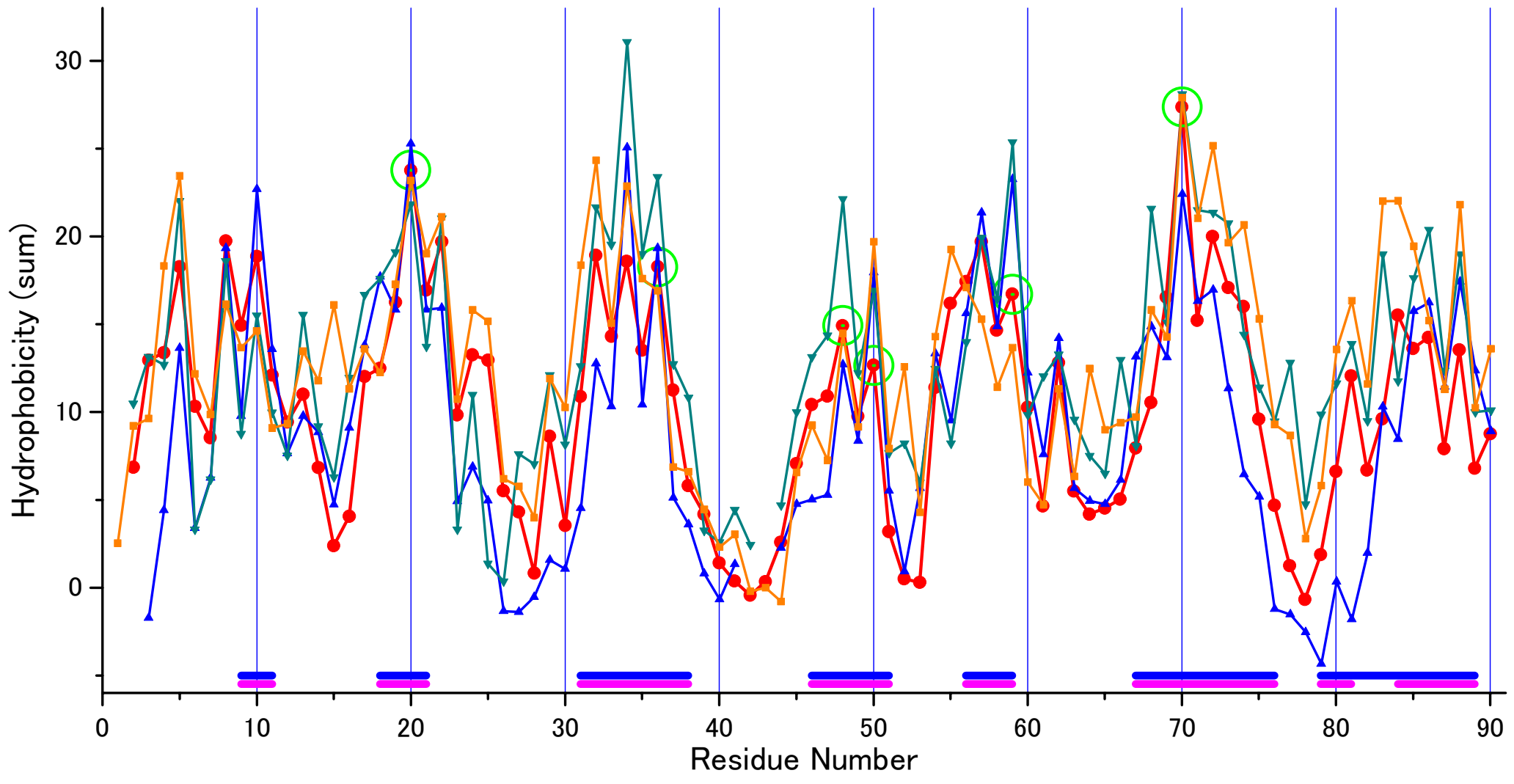

*Figure 4*. The 3D hydrophobicity profiles for 1ten (●) and 3 proteins, 1qr4A (▲), 1fnhA (▼) and 2mfn (■), with the highest *Z*-scores, *Z*>14.5, in HSSP. The protein 1fnf with *Z*=16.6 is not included in this analysis, since it has many gapped regions in sequence alignment.

In this section we have found that the 3D-HP is essential to specify the structure in the native state and evolutionally conserved. This 3D-HP is a mean-field description of interactions, including entropic effect, among amino acids.

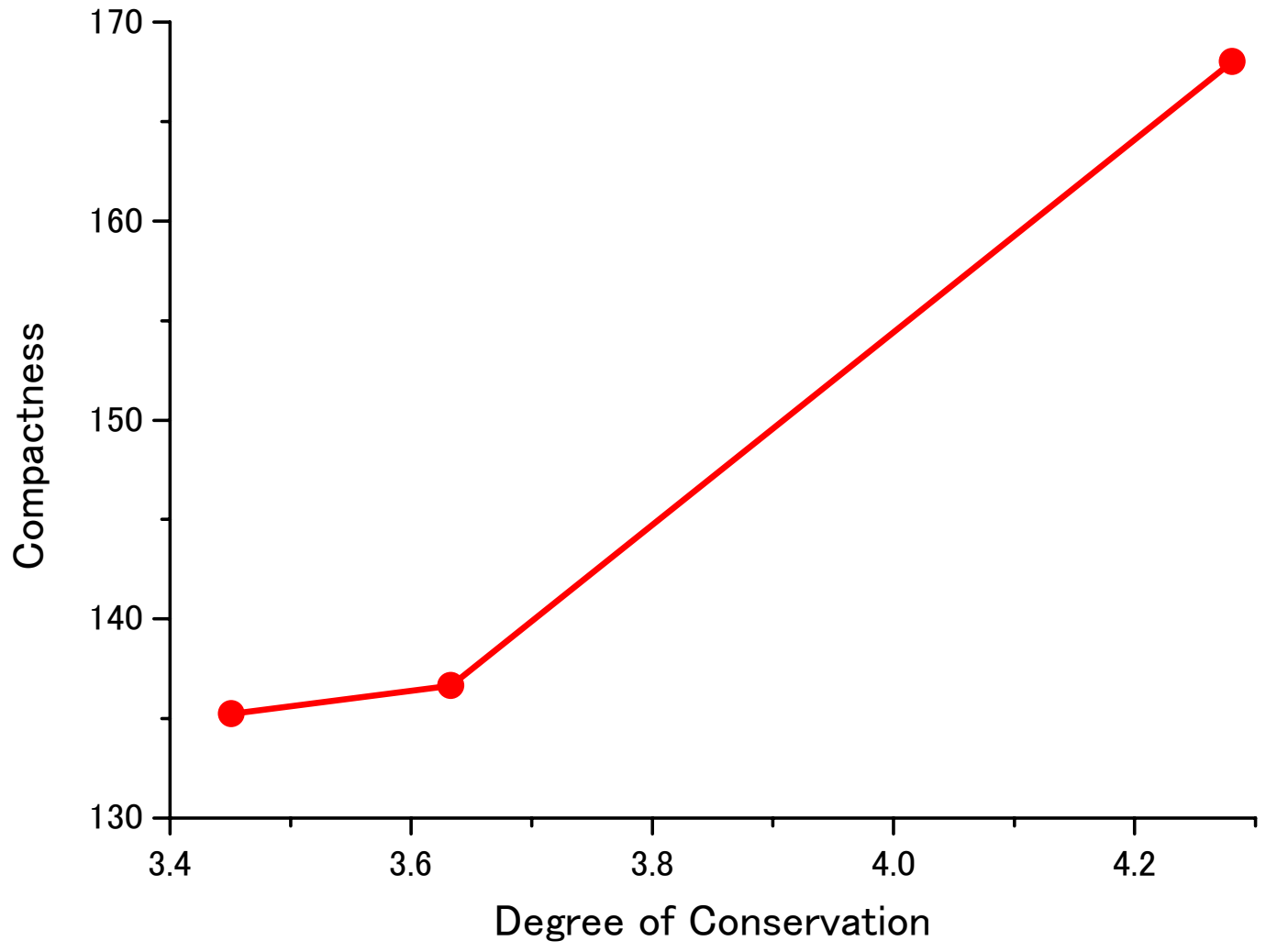

*Figure 5*. The correlation between the degree of conservation and the compactness. The degree of conservation is the mean of the deviation squared of the 3D hydrophobicity subtracted the mean 3D hydrophobicity over the whole sequence, for each protein. The compactness is the sum of the distances among 6 folding nuclei. The folding nuclei for 3 proteins, 1qr4A, 1fnhA and 2mfn, are assumed to be the same as 1ten in residue number. The compactness for 1ten is 130.94.

## 4. $\phi$ -value and Contact Distance

In the previous section we have shown that the 3DHP is evolutionally conserved. However, the correlation between the 3DHP and the $\phi$ -value at the folding nuclei is not so good. We see a tendency that the larger $\phi$ -values are observed at the middle part of the amino-acid chain: (the residue number; the $\phi$ -value, the 3D hydrophobicity): (20; 0.38, 23.77), (36; 0.56, 18.27), (48; 0.67, 14.92), (50; 0.42, 12.66), (59; 0.62, 16.71) and (70; 0.54, 27.37). Thus we consider the effect of the contact distance. In Figure 6 the correlation between the ratio, the $\phi$ -value versus the 3D hydrophobicity, and the contact distance is shown. This correlation reflects the fact that the fast processes of the folding correspond to the formation of native contacts at short distance in 1D amino-acid sequence and the slow processes at long distance.

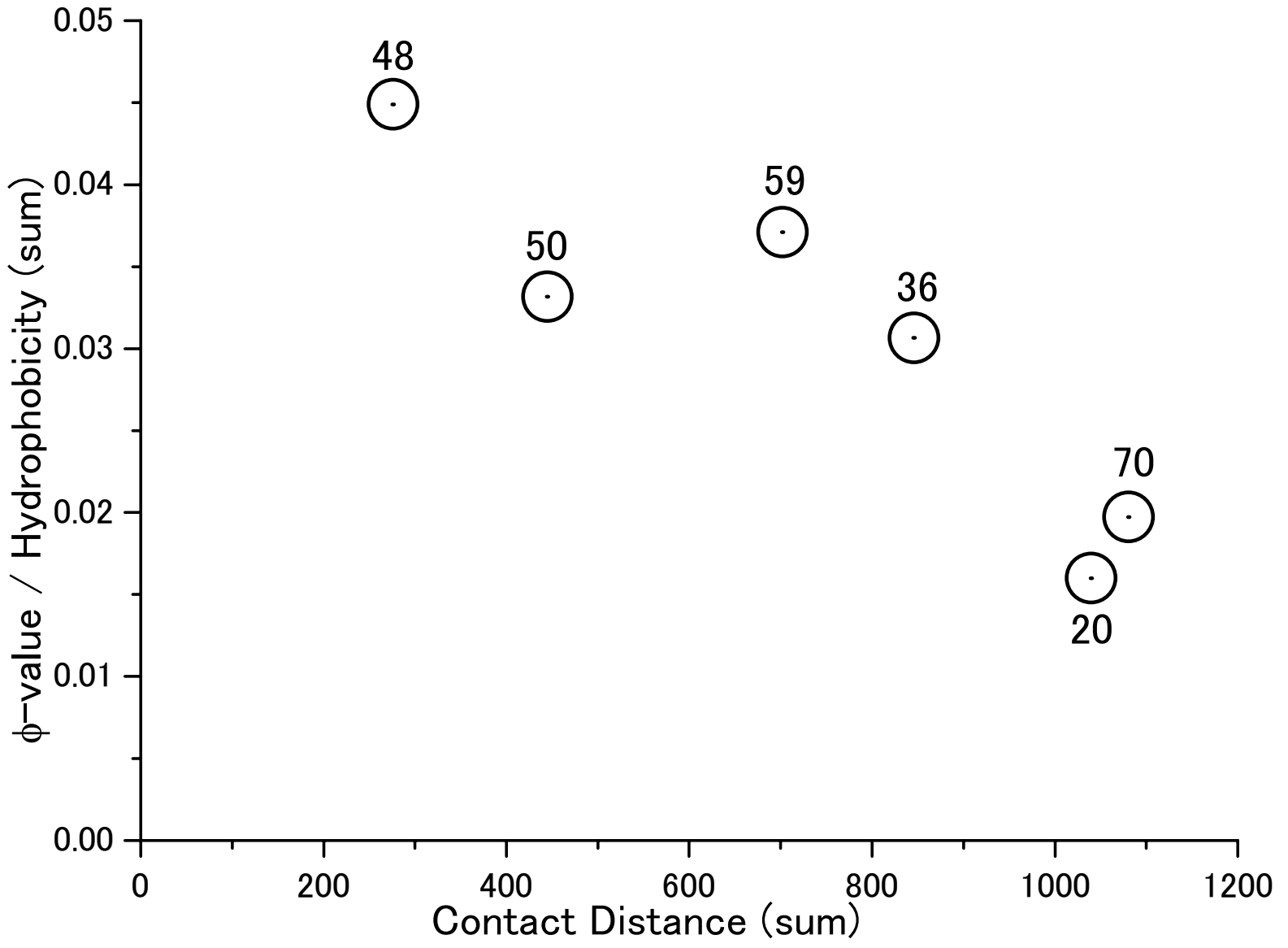

*Figure 6.* The correlation between the ratio, the $\phi$ -value / the 3D hydrophobicity, and the sum of the contact distance for folding nuclei. The contact distance is the difference in residue numbers between two amino acids contacting in the native state structure. The condition for contacts is the same as that employed in Figure 3. A similar residue-wise contact order has been discussed in another context [9].

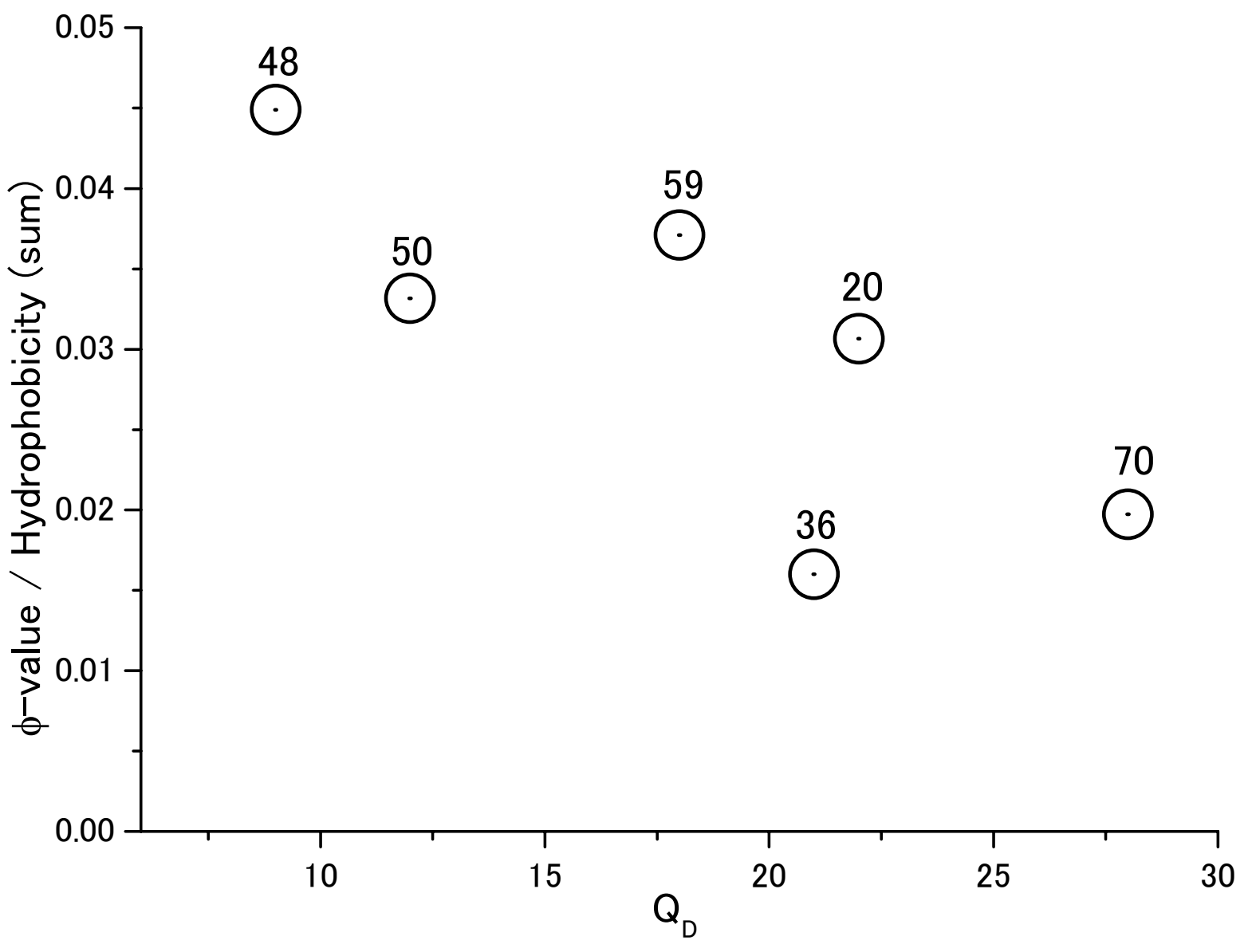

*Figure 7*. The correlation between the ratio, the $\phi$-value / the 3D hydrophobicity, and the number of native pairs $Q_D$ for folding nuclei. A native pair is a pair of two amino acids which are within 12 Å in the native state structure and whose distance in residue number is larger than 12. The distance between amino acids is determined by that for $C_\alpha$'s. Original $Q_D$ has been discussed in a review [10].

The same analysis can be done using the number of native pairs, $Q_D$, instead of the contact distance. In Figure 7 the correlation between the ratio, the $\phi$-value versus the 3D hydrophobicity and $Q_D$ is shown. This correlation is weaker than that in Figure 6. Such a result supports the importance of the contact distance, since $Q_D$ lacks the information of the distance in 1D amino-acid sequence.

## 5. Conclusion and Discussions

In conclusion we have shown that the 3DHP is evolutionally conserved and specifies the folding nuclei. The 3DHP correlates with experimentally measured $\phi$-value after correcting the effect of the contact distance. Since the 3DHP is a physico-chemical property averaged over the 3D structure of the native state, the information of the 1D amino-acid sequence is rather unimportant. This fact can be an explanation of the experimental evidence that the 3D structure is evolutionally more conserved than the 1D sequence, since the same 3DHP can be realized by plural different 1D sequences.

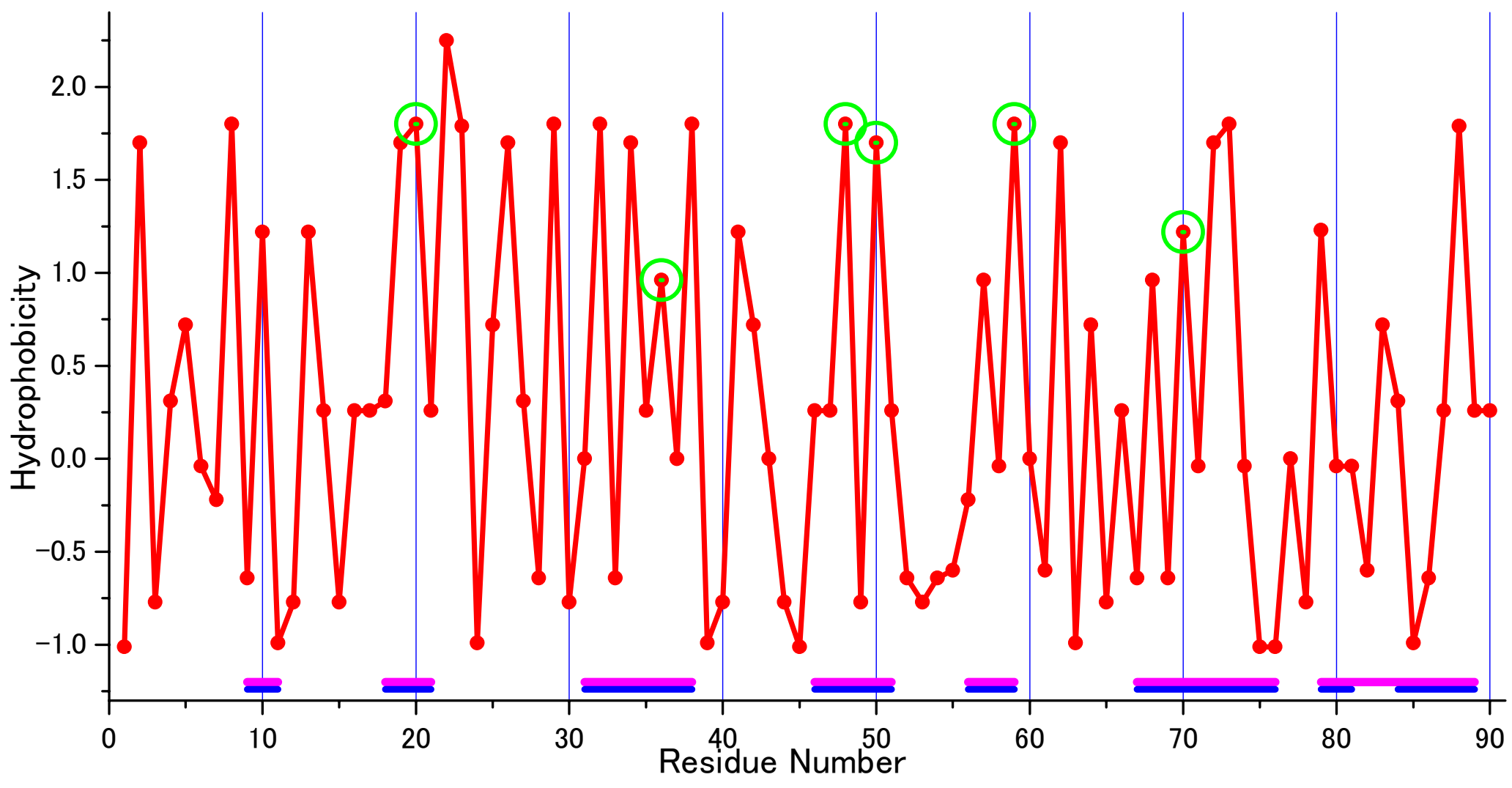

*Figure 8*. The hydrophobicity for each amino acid as a function of sequence.

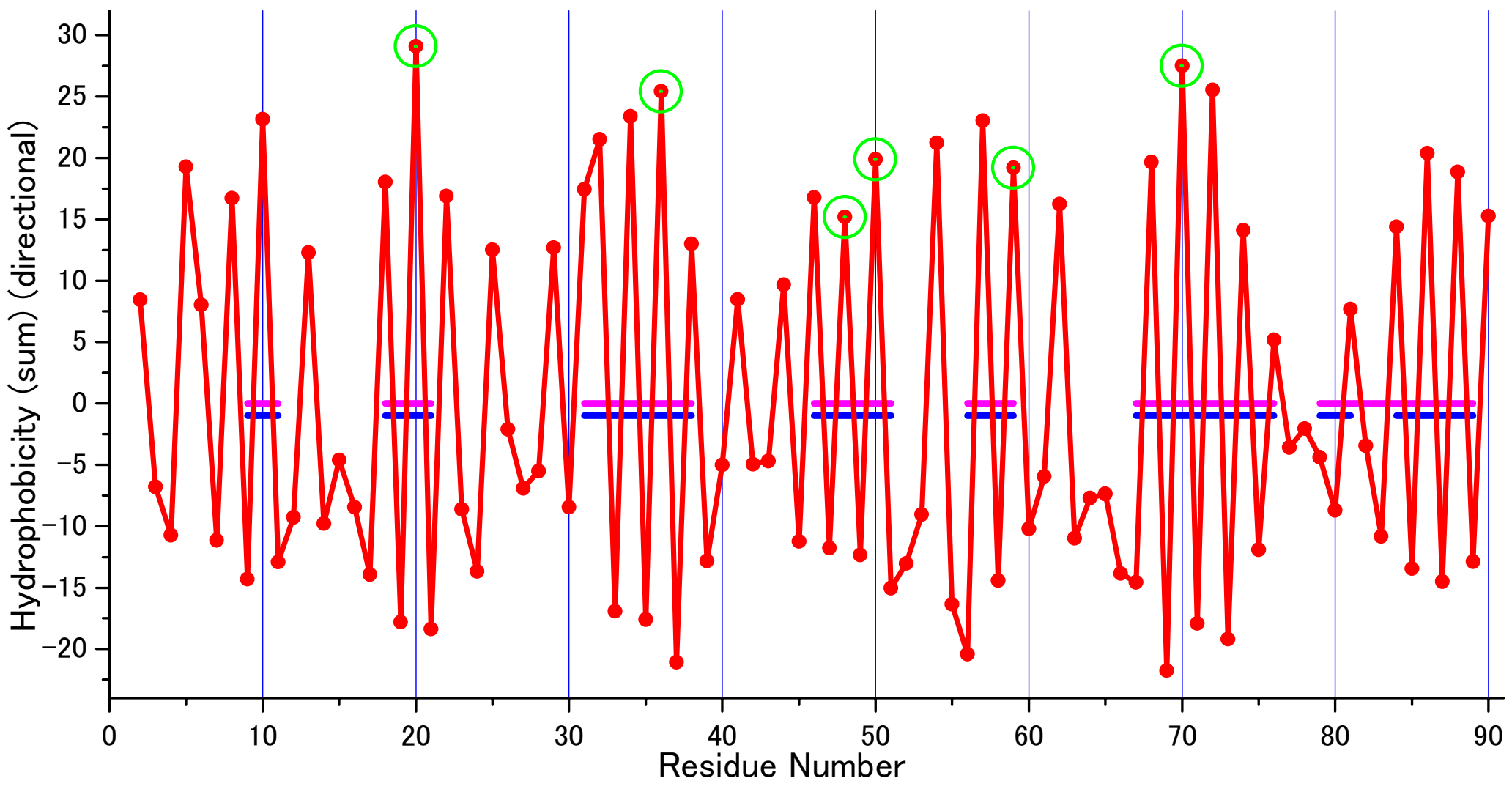

*Figure 9*. The 3D hydrophobicity profile with the directional weight as a function of amino-acid sequence. The 3D hydrophobicity for an amino acid in this case is defined in a similar manner as in Figure 3 but in the calculation of the sum the directional weight is multiplied where the weight is +1 for the amino acids whose side chain is in the inside of 1ten and –1 for outside.

In our study it is clarified that the mean value of a physico-chemical property, the 3DHP, is optimized and evolutionally conserved. Figures 8 and 9 also support the significance of the mean value. In Figures 8 and 9 it is demonstrated that the zigzag pattern in hydrophobicity along the amino-acid sequence, which leads to $\beta$-sheet structure, becomes more evident in the mean value than in the original value of the amino acid. Thus the property of the amino acid itself is rather unimportant but the mean value realized in the native state structure after interacting with the other amino acids and surrounding water molecules.

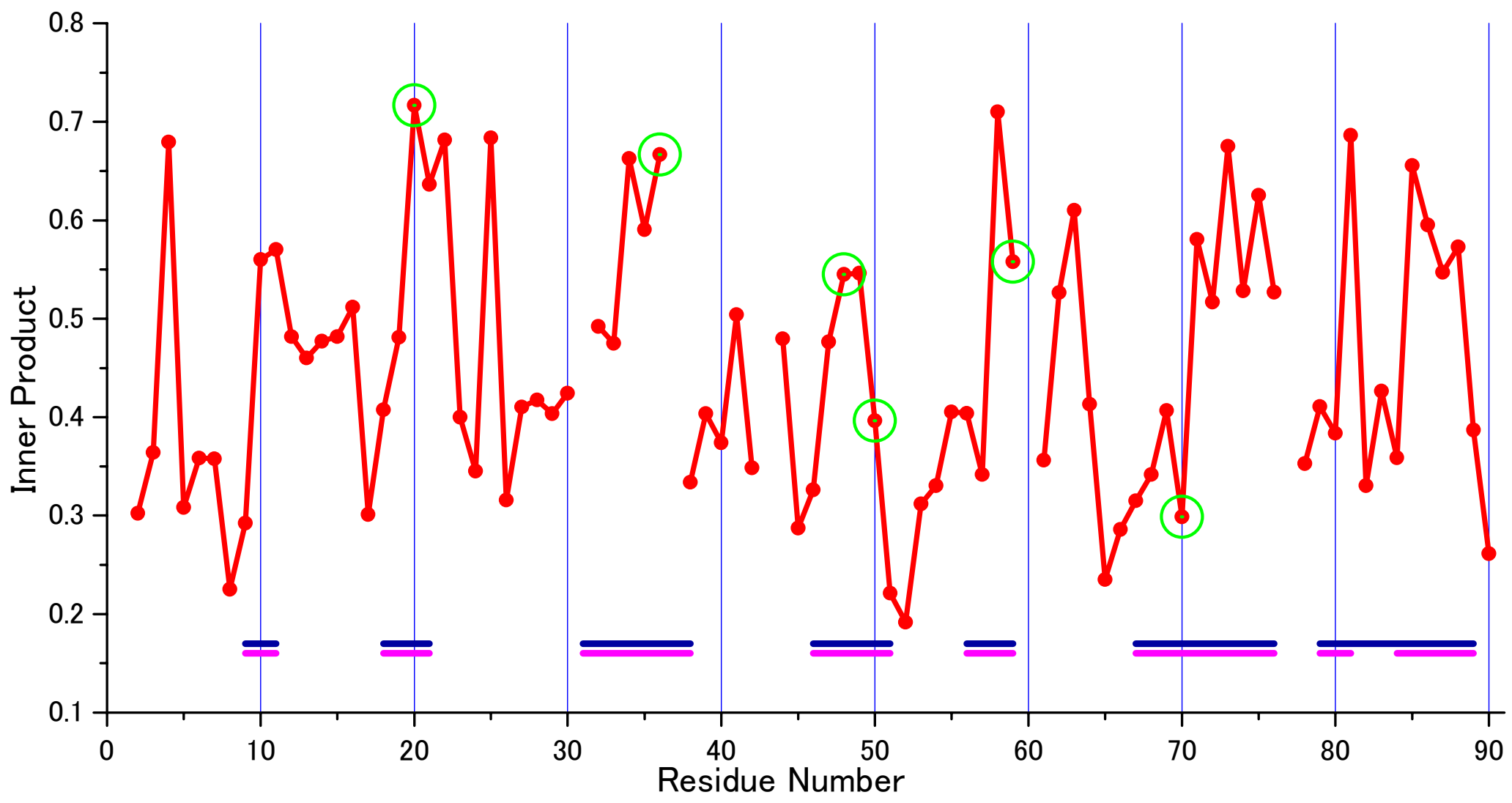

*Figure 10*. The degree of distortion of the side chain, the inner product, in the native state structure as a function of amino acid sequence. The inner product is calculated for two normalized vectors, one for the direction of the side chain determined by $C_\beta$ in AMBER and the other in the native state structure.

In Figure 10 we show the difference in the directions of the side chain between the amino acid in the native state and that in the isolated state. It is found that the difference is rather small at the folding nuclei. On the other hand, in order to form tightly packed hydrophobic core, the side chains of amino acids involved into the core at later stage, inevitably deformed.

In our study we have focused our attention to the folding nuclei on secondary structures in consistent with the *S*-value analysis [11]. Here we have assumed a hierarchical picture of the folding where the weak residual interactions determine the 3D structure among secondary structures and the stronger interactions, for example hydrogen bonding, lead to secondary structures before the process of 3D structure formation.

In this paper we have done a case study for 1ten, since it is simple and thoroughly studied mini-protein. In future the same analysis done in this paper should be applied to other proteins and our scenario should be tested for wide class of proteins.

## Appendix

In the first version of `physics/0607219`, we have proposed a field theory of 3DHP that explain $\phi$-value analysis for protein folding. In this Appendix we illustrate the essence of the field theory. This is a key concept to understand the correspondence between sequence and structure of proteins.

The 3DHP is a field variable in the native state. We have shown that such a field variable is more important than the sequence information of amino acids. Thus the structure of proteins is evolutionally conserved even if the sequence is not conserved. The field pattern is determined by the interaction among amino acids and a specific pattern can be realized by plural different sequences. This is illustrated in the following figure.

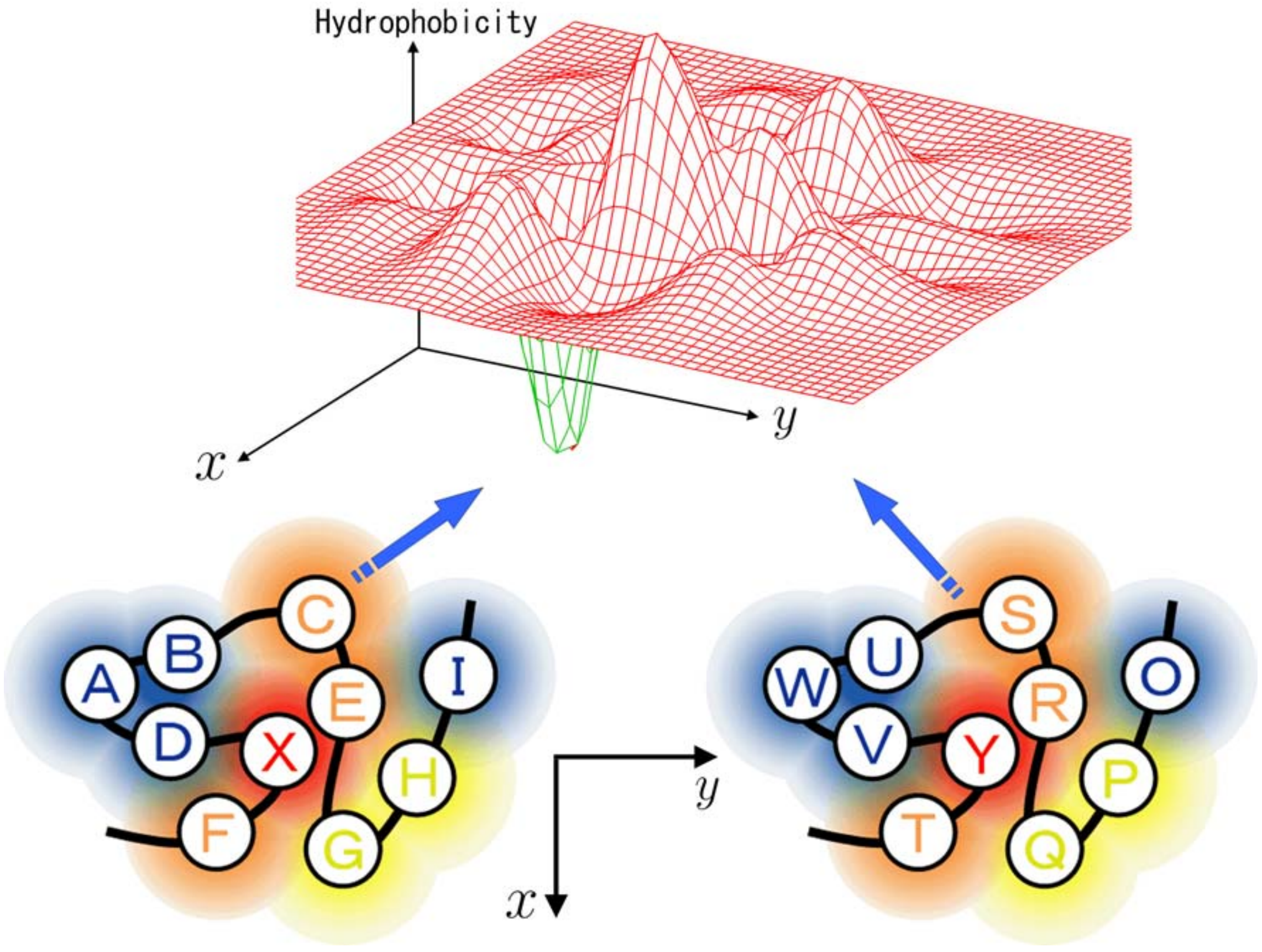